\documentclass[letterpaper, 10 pt, conference]{ieeeconf}
\IEEEoverridecommandlockouts 
\usepackage[T1]{fontenc}
\usepackage{caption,color}
\usepackage{graphicx}
\usepackage{amsmath,amssymb,amsfonts}
\usepackage{textcomp}
\usepackage{dsfont}
\usepackage{bbold}
\usepackage{arydshln}
\usepackage{comment}
\usepackage{mathtools} 
\usepackage{mathrsfs}
\usepackage{enumerate}
\everymath{\displaystyle}
\usepackage{cases}
\usepackage[framed,thmmarks]{ntheorem}  
\usepackage[overload]{empheq}
\usepackage{booktabs}
\usepackage{datetime}
\usepackage{cite}
\usepackage{stfloats}
\usepackage{pifont}
\newenvironment{proofa}{{\noindent\bfseries{Proof:}}}{\hfill$\blacksquare$}
\usepackage{etoolbox}

\newtheorem{lemma}{Lemma}
\newtheorem{assumption}{Assumption}
\newtheorem{remark}{Remark}
\newtheorem{definition}{Definition}
\AtEndEnvironment{problem}{\hfill$\Box$}
\AtEndEnvironment{lemma}{\hfill$\Box$}
\AtEndEnvironment{theorem}{\hfill$\Box$}
\AtEndEnvironment{assumption}{\hfill$\Box$}
\AtEndEnvironment{condition}{\hfill$\Box$}
\AtEndEnvironment{remark}{\hfill$\Box$}
\AtEndEnvironment{corollary}{\hfill$\Box$}
\AtEndEnvironment{stassumption}{\hfill$\Box$}
\AtEndEnvironment{definition}{\hfill$\Box$}

\title{\textbf{State Estimation Using Single Body-Frame
Bearing Measurements}}

\def\namedlabel#1#2{\begingroup
    #2%
    \def\@currentlabel{#2}%
    \phantomsection\label{#1}\endgroup
}
\author{Sifeddine Benahmed, Soulaimane Berkane, \IEEEmembership{Senior Member, IEEE}
\thanks{Sifeddine Benahmed is with Capgemini Engineering, Toulouse, 31300, France (email: sif-eddine.benahmed@capgemini.com).}
\thanks{Soulaimane Berkane is with the Department Computer Science and Engineering, University of Quebec in Outaouais, QC J8X 3X7, Canada (e-mail:
soulaimane.berkane@uqo.ca).}
\thanks{* This research work is supported in part by NSERC-DG RGPIN-2020-04759 and the Fonds de recherche du Qu\'ebec (FRQ).}
\thanks{}}

\theoremstyle{dotlessP}

\newcommand{\RNum}[1]{\uppercase\expandafter{\romannumeral #1\relax}}

\newcommand{\R}{\mathds{R}}

\newcommand{\sphere}{\mathds{S}}

\newcommand{\iframe}{\{\mathcal{I}\}}
\newcommand{\bframe}{\{\mathcal{B}\}}
\newcommand{\so}{\mathds{SO}(3)}
\newcommand{\wx}{[\omega]_{\times}}
\newcommand{\Rthree}{\R^{3}}
\newcommand{\liealg}{\mathbf{so}(3)}
\newcommand{\posif}{p^{\mathcal{I}}}
\newcommand{\dposif}{\dot{p}^{\mathcal{I}}}
\newcommand{\velif}{v^{\mathcal{I}}}
\newcommand{\dvelif}{\dot{v}^{\mathcal{I}}}
\newcommand{\landposif}{p_{\ell}^{\mathcal{I}}}
\newcommand{\othree}{0_{3\times 3}}
\newcommand{\gravbfr}{g^{\mathcal{B}}}
\newcommand{\hatgravbfr}{\hat{g}^{\mathcal{B}}}
\newcommand{\gravifr}{g^{\mathcal{I}}}
\newcommand{\posbfr}{p^{\mathcal{B}}}
\newcommand{\velbfr}{v^{\mathcal{B}}}
\newcommand{\accelbfr}{a^{\mathcal{B}}}
\newcommand{\magbfr}{m^{\mathcal{B}}}
\newcommand{\hatmagbfr}{\hat{m}^{\mathcal{B}}}
\newcommand{\magifr}{m^{\mathcal{I}}}
\newcommand{\bearingbfr}{\eta^{\mathcal{B}}}
\newcommand{\bearingifr}{\eta^{\mathcal{I}}}
\begin{document}
\maketitle 


\begin{abstract}
This paper addresses the problem of simultaneous estimation of the position, linear velocity and  orientation of a rigid body using single bearing measurements. We introduce a Riccati observer-based estimator that fuses measurements from a 3-axis accelerometer, a 3-axis gyroscope, a single body-frame vector observation (\textit{e.g.,} magnetometer), and a single bearing-to-landmark measurement to obtain the full vehicle's state (position, velocity, orientation). The proposed observer guarantees global exponential convergence under some persistency of excitation (PE) condition on the vehicle's motion. Simulation results are presented to show the effectiveness of the proposed approach.
\end{abstract}
\section{Introduction}
The problem of accurate complete pose (position, velocity, and orientation) estimation of a rigid body is a critical challenge for autonomous robotic platforms \cite{Titterton2004StrapdownTechnology}. Inertial navigation systems (INS) serve as essential devices, in this context, enabling the localization and control of autonomous vehicles by fusing measurements from onboard sensors. These sensors are essentially accelerometers and gyroscopes (typically included in an Inertial Measurement Unit (IMU)). INS computes the position, velocity, and orientation by direct integration of the information provided by these sensors. However, this approach can easily fail in case of measurements errors or unknown initial conditions \cite{Woodman_INS_tech_report}. Therefore, INS requires often additional sensors such as Global Positioning System (GPS) to correct the position estimates over time. However, in GPS-denied environments, such as indoor applications, other types of sensors are required. Vision and acoustic sensors \cite{Marco_ECC_2020,reis2018source}, among others, can be used to provide bearing (direction) measurements. These measurements can also be combined with body-frame vector measurements obtained from onboard sensors, such as magnetometer and dual-GPS, to achieve accurate pose estimation.

Numerous estimation algorithms have been developed to enhance INS through the integration of bearing measurements and body-frame vector observations. These algorithms find several applications in a wide array of fields, including source localization \cite{Old0source0Reed0Jesse}, multiple object tracking \cite{multiple0trakingKim2022}, marine navigation \cite{EKF0Marina0application0Zhao}, and cooperative navigation \cite{cooperative0navigation0SANTOS2020}. Most of these techniques are of Kalman-type. Despite their recognition as industry-standard solutions, these stochastic filters come with inherent limitations. One notable drawback is their dependence on local linearization, making them vulnerable to initialization errors. To overcome these limitations, several nonlinear deterministic observers have been developed 
recently. These observers offer distinct advantages, including well-established stability guarantees and computational simplicity in contrast to 
stochastic filters, as mentioned in \cite{Berkane_Automatica_2021}. 

In the context of vision-aided INS, a nonlinear observer relying on 3D position landmark measurements with global exponential convergence is proposed in \cite{wang2020hybrid}. To overcome the need for landmarks 3D positions reconstruction \cite{hartley2003multiple}, a nonlinear observer is presented in \cite{wang2019nonlinear}  with direct stereo bearing (direction) measurements, while also providing global stability guarantees under specific conditions on the number and the configuration of the landmarks. On the other hand, monocular cameras offer advantages in terms of cost, simplicity, and reduced weight compared to stereo-cameras. Therefore, in \cite{marco2020position}, a local Riccati observer for simultaneous estimation of attitude, position, linear velocity, and accelerometer bias with monocular-bearing measurements is proposed, however it achieves only local exponential convergence. In the context of underwater applications, global exponential convergent observers are proposed in \cite{reis2018source, batista2013globally} relying on bearing measurements from acoustic sensors and relative velocity information.

In this paper, we propose a Riccati observer-based estimation scheme for state estimation using IMU and single bearing measurements. In contrast to \cite{Hamel2017PositionMeasurements,reis2018source,batista2013globally}, and with an additional body-frame vector observation ({\it e.g.,} obtained from a magnetometer), our estimator provides complete estimation of the vehicle's inertial position, inertial velocity, and orientation. In fact, in this work, we do not make use of the low-acceleration assumption commonly used to decouple the problem of IMU-based attitude estimation and position estimation, see \cite{berkane2021nonlinear,Berkane_Automatica_2021} for a motivation.  Under a persistency of excitation (PE) condition on the vehicle's motion relative to the landmark, the proposed observer guarantees global exponential stability; a strong stability result which cannot be obtained using geometric observers such as \cite{Wang_TAC_2022} due to topological obstructions. Furthermore, using IMU and single bearing only, a reduced-order version of the observer allows to estimate position, velocity, and gravity direction in body-frame. The body-frame gravity direction allows, for example, to extract reduced attitude (roll and pitch). In this case, yaw (heading) readings obtained independently from other external sources (\textit{e.g.,} dual GPS, compass, etc) allow to recover the full vehicle's state.

The rest of the paper is organized as follows. The next section provides some preliminaries. Section~\ref{Section:Problem_formulation} formulates the considered problem where details about the studied vehicle’s model, the possible available measurements and
the technical assumptions needed for our main result are provided. In Section~\ref{Section:main_result}, we present the main result with the proposed observer and the associated observability analysis. In Section~\ref{section:simulation}, we provide simulation results while Section~\ref{section:conclusion} concludes the paper.

\section{Preliminaries}\label{section:Notation_Preliminaries}
We denote by $\mathds{Z}_{>0}$  the set of positive integers, by $\R$ the set of reals, by $\R^n$ the $n$-dimensional
Euclidean space, and by $\sphere^n$ the unit $n$-sphere embedded in
$\R^{n+1}$. We use $||x||$ to denote the Euclidean
norm of a vector $x\in\R^n$. The $i$-th element of a vector $x\in\R^{n}$ is denoted by $x_i$. The $n$-by-$n$ identity and zeros matrices are denoted by $I_n$ and $\othree$, respectively. By $\mathrm{blkdiag}(\cdot)$, we denote the
block diagonal matrix. The Special Orthogonal group of order three is denoted
by $\so:= \{A\in\R^{3\times3}: \mathrm{det}(A) = 1; AA^{\top} =A^{\top}A=I_3\}$. The set 
$\liealg:=\{\Omega\in\R^{3\times3}:\Omega=-\Omega^{\top}\}$ denotes the Lie algebra of $\so$. For $x,\ y\in\Rthree$, the map $[.]_{\times}:\Rthree\to\liealg$ is defined such that $[x]_{\times}y=x\times y$ where $\times$ is the vector cross-product in $\Rthree$. We introduce the following important orthogonal projection operator $\Pi:\sphere^2 \to \Rthree$
that will be used throughout the paper:
\begin{equation}
\Pi_x=I_3-xx^{\top},\quad x\in\sphere^2.
\end{equation}
Note that $\Pi_x$ is an orthogonal projection matrix which geometrically
projects any vector in $\Rthree$ onto the plan orthogonal to vector $x\in\sphere^2$.
In addition, one verifies that $\Pi_xy=0_{3\times1}$ if $x$ and $y$ are collinear. For simplicity and for the sake of clarity, the argument of the time-dependent signals is omitted unless otherwise required.

\subsection{State Estimation for Linear Time-Varying (LTV) Systems}
An LTV system is described by 
\begin{subequations}\label{equation:general_LTV_state_model}
    \begin{align}
    \dot{x}&=A(t)x+B(t)u,\\
    y&=C(t)x,
    \end{align}
\end{subequations}
where $x\in\R^{n}$ is the state, $u\in\R^{m}$  is the input and $y\in\R^p$  is the output with $n,m,p\in\mathds{Z}_{>0}$. The time-varying matrices $A(t)\in\R^{n\times n}$, $B(t)\in\R^{n \times m}$, $C(t)\in\R^{p \times n}$ are known, continuously differentiable and uniformly bounded with bounded derivatives.
Observer design for LTV systems has a long history. Mainly, in the spirit of the Kalman filter, solutions are usually based on a Luenberger-type observer with a gain matrix updated using some sort of a Riccati equation \cite{besanccon2007overview,Hamel2017PositionMeasurements}. A traditional Riccati observer for system~\eqref{equation:general_LTV_state_model} is given by
\begin{align}\label{equation:Riccati_observer}
\dot{\hat{x}}&=A(t)\hat{x}+B(t)u+K(t)(y-C(t)\hat{x}),
\end{align}
with $\hat{x}$ is the estimate of $x$, and the gain of the observer is given by
\begin{equation}\label{equation:gain_general_Riccati_observer}
  K(t)=PC(t)^{\top}Q(t),  
\end{equation}
where $P$ is the solution of the following Riccati equation:
\begin{equation}\label{equation:Riccati_equation}
    \dot{P}=A(t)P+PA^{\top}(t)-PC^{\top}(t)Q(t)C(t)P+V(t),
\end{equation}
and where $P(0)$ is a positive definite matrix and  $Q(t)$ and $V(t)$ are uniformly positive definite matrices that should be specified. Note that, in the context of Kalman filter, the matrices $V(t)$ and $Q^{-1}(t)$ represent covariance matrices characterizing additive noise on the system state. 

The following definition formulates the well-known uniform observability (UO) condition in terms of the observability Gramian matrix. The UO property guarantees uniform global exponential stability of the Riccati observer  \eqref{equation:Riccati_observer}, see \cite{Hamel2017PositionMeasurements} for more details.

\begin{definition}\textbf{(Uniform Observability)}\label{definition:uniform_observability} The pair $(A(t)$$,C(t))$ is uniformly  observable if there exist constants $\delta,\mu>0$ such that
\begin{align}\label{equation:conditon_of_uniform_observability}
    W(t,t+\delta)&:=\frac{1}{\delta}\int_t^{t+\delta}\phi^{\top}(s,t)C^{\top}(s)C(s)\phi(s,t)ds \nonumber\\
    &\geq\mu I_n, \quad  \forall t\geq0
\end{align}
where $\phi(s,t)$ is the transition matrix associted to $A(t)\in\R^{n\times n}$ such that $\frac{d}{dt}\phi(t,s)=A(t)\phi(t,s)$ and $\phi(t,t)=I_n$.
\end{definition}

\section{Problem Formulation}\label{Section:Problem_formulation}
Let $\iframe$ be an inertial frame, $\bframe$ be an NED body-fixed frame attached to the center of mass of a rigid body (vehicle) and the rotation matrix $R\in\so$ be the orientation (attitude) of frame $\bframe$ with respect to $\iframe$. Consider the following 3D kinematics of a rigid body
\begin{subequations}\label{equation:dynamic_model_Inertial_frame}
\begin{align}
\label{eq:dp}
\dposif&=\velif,\\
\label{eq:dv}
\dvelif&=\gravifr+R\accelbfr,\\
\dot{R}&=R\wx,
\end{align}
\end{subequations}
where the vectors $\posif\in\R^{3}$ and $\velif\in\R^{3}$ denote the position and linear velocity of the rigid body expressed in frame $\iframe$, respectively, $\omega$ is the angular velocity of $\bframe$ with respect to $\iframe$ expressed in $\bframe$, $\gravifr\in\R^{3}$ is the gravity vector expressed in $\iframe$, and  $\accelbfr\in\R^{3}$ is the 'apparent acceleration' capturing all non-gravitational forces applied to the rigid body expressed in frame $\bframe$.

This work focuses on the problem of position, linear velocity and attitude estimation for INS. 
The objective of this paper is to design a uniformly globally  convergent observer to simultaneously estimate the inertial position $\posif$, inertial velocity $\velif$ and attitude $R$ using the following measurements:
\begin{assumption}[Available Measurements]\label{assumption:available_measurements}
We assume that the following measurements are available:\\
$\quad (i)$ The angular velocity $\omega$.\\
$\quad (ii)$ The apparent acceleration $\accelbfr$.\\
$\quad (iii)$ A body-frame bearing measurement to a known landmark.\\
$\quad (iv)$ A body-frame vector measurement of a known inertial direction.
\end{assumption}

The measurements in items $(i)$ and $(ii)$ of Assumption~\ref{assumption:available_measurements} can be obtained from an IMU while the one of item $(iii)$ of Assumption~\ref{assumption:available_measurements} can be obtained, for instance, from vision or acoustic sensors. The single bearing measurement to the landmark expressed in $\bframe$ is given by
\begin{equation}\label{equation:bearing_measurement}
\bearingbfr:=R^{\top}\frac{p^{\mathcal{I}}-\landposif}{||p^{\mathcal{I}}-\landposif||},
\end{equation}
where $\landposif\in\Rthree$ is the position (constant and  known) of the landmark in $\iframe$, see Fig.~\ref{figure:illustration_of_bearing_vector}. Note that the measurement $\bearingbfr$ gives information solely about the direction to the landmark with respect to $\bframe$. More specifically, the unitary vector $\bearingbfr$ corresponds to the projection of the landmark position vector with respect to $\bframe$ onto a virtual spherical image plane.
The body-frame vector measurement in item~$(iv)$ of Assumption~\ref{assumption:available_measurements} can be obtained using an additional sensor, \textit{e.g.,} a magnetometer. This vector correspond to the expression in $\bframe$ of a constant and known vector $\magifr\in\R^{3}$ in $\iframe$, {\it i.e.,}
\begin{equation}\label{equation:magnetic_filed}
    \magbfr=R^{\top}\magifr.
\end{equation}
The following is a general observability assumption used in the field of attitude estimation (see, \textit{e.g.,}  \cite{mahony2008nonlinear}).
\begin{assumption}\label{assumption::obsv}
The inertial vectors $\magifr$ and $\gravifr$ are noncollinear. 
\end{assumption}
\begin{figure}[h!]
    \centering    \includegraphics{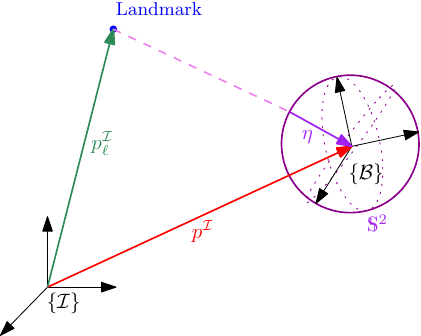}
    \caption{Illustration of the bearing measurement $\bearingbfr$, the position of the rigid body $p^{\mathcal{I}}$ and the landmark $\landposif$.}
    \label{figure:illustration_of_bearing_vector}
\end{figure}
Note that Assumption \ref{assumption:available_measurements} provides a very minimal set of measurements for the given problem. In fact, for the attitude estimation problem alone, we usually require at least two non-collinear body-frame vector observations \cite{mahony2008nonlinear} while the translational motion estimation usually requires position information in inertial frame as in \cite{Hamel2017PositionMeasurements,batista2013globally}. In this work, we use only a single vector observation along with a body-frame position bearing to recover position, velocity, and orientation.

The translational system \eqref{eq:dp}-\eqref{eq:dv} is a linear system with an unknown input $R\accelbfr$. Therefore, there is a coupling between the translational dynamics and the rotational dynamics through the accelerometer measurements. Most adhoc methods in practice assume that $R\accelbfr\approx -\gravifr$ to remove this coupling between the translational and rotational dynamics. However, this assumption holds only for non-accelerated vehicles, {\it i.e., } when $\dvelif\approx 0$. In this work, we instead design our estimation algorithm without this latter assumption.

\section{Main Results}\label{Section:main_result}
In this section, we provide the main result of this work. 
We first write the state-space model of the system in the body frame as well as a virtual output defined by the projection of the rigid body's position (in $\bframe$) in the plan perpendicular to the bearing measurement $\bearingbfr$. We then establish a sufficient condition for uniform observability of the resulting LTV system.

Let $\posbfr=R^{\top}\posif$, $\velbfr=R^{\top}\velif$, $\gravbfr=R^{\top}\gravifr$ be the position and linear velocity of the rigid body and the gravity vector, all expressed in $\bframe$, respectively. Thus, in view of \eqref{equation:dynamic_model_Inertial_frame} and \eqref{equation:magnetic_filed}, we have:
\begin{subequations}\label{equation:dynamic_model_Body_frame}
 \begin{align}
\dot{p}^{\mathcal{B}}&=-\wx \posbfr+\velbfr,\\
\dot{v}^{\mathcal{B}}&=-\wx \velbfr + \accelbfr+\gravbfr,\\
\dot{g}^{\mathcal{B}}&=-\wx \gravbfr,\\
\dot{m}^{\mathcal{B}}&=-\wx \magbfr.
\end{align}
\end{subequations}

To simplify the analysis, we assume without loss of generality that $\landposif=0$, \textit{i.e.,} the center of $\iframe$ coincides with the position of the landmark. Note that if this was not the case, one can redefine $\tilde{p}^{\mathcal{I}}=\posif-\landposif$ and $\dot{\tilde{p}}^{\mathcal{I}}=\velif$. Once $\tilde{p}^{\mathcal{I}}$ is estimated then $\posif=\tilde{p}^{\mathcal{I}}+\landposif$. Hence, \eqref{equation:bearing_measurement} becomes 
\begin{equation}\label{equation:bearing_measurement_simplified}
\bearingbfr=R^{\top}\frac{p^{\mathcal{I}}}{||p^{\mathcal{I}}||}=\frac{\posbfr}{||\posbfr||}.
\end{equation}
Let $\tilde{y}:=\Pi_{\bearingbfr} \posbfr$. In view of \eqref{equation:bearing_measurement_simplified}, we have $\bearingbfr$ and $\posbfr$ are collinear and thus
\begin{equation}\label{equation:virtual_nul_output}
\tilde{y}=0.
\end{equation}
Equation \eqref{equation:virtual_nul_output} represents a virtual (linear in $\posbfr$) output inspired from \cite{Batista2015NavigationMeasurements,Hamel2017PositionMeasurements,Berkane_Automatica_2021}. This form of the output will be useful in the design of the observer as well as in conducting the corresponding uniform observability analysis.
\begin{figure*}[h!]
    \centering    \includegraphics[scale=0.8]{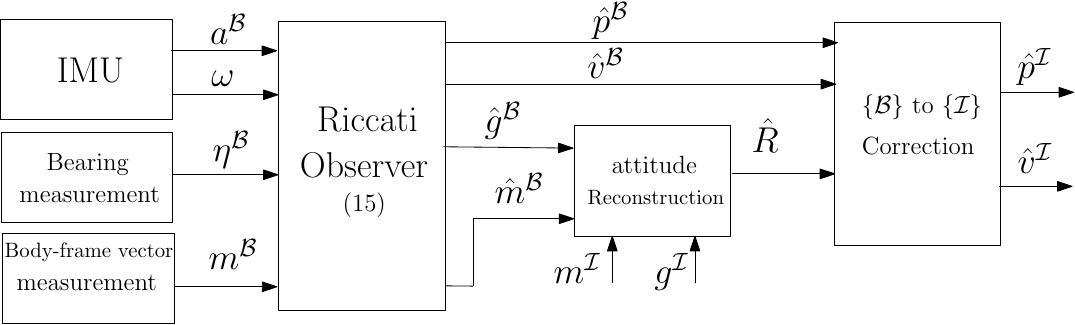}
    \caption{Illustration of the proposed state estimation approach.}
    \label{figure:Illustration_of_the_proposed_method}
\end{figure*}
 The structure of the proposed approach is given in Fig.~\ref{figure:Illustration_of_the_proposed_method}. The proposed observer provides an estimation of $\posbfr$, $\velbfr$, $\gravbfr$ and $\magbfr$ using the available measurements. The estimates of $\gravbfr$ and $\magbfr$ together with their expressions in $\iframe$ are then fed to an attitude reconstruction algorithm.  Once the attitude is reconstructed, the estimates of the position and the velocity in the inertial frame are recovered.

Note that $\gravbfr$ is also estimated because, for accelerated vehicles, the accelerometer does not measure the body-frame gravity vector (measures only non-gravitational forces). Besides, it should be mentioned that it is possible to obviate $\magbfr$ in the estimator's state but this measurement has been included for noise filtering purposes, see also Remark \ref{remark2} for an alternative reduced-order estimator. 

Now, by letting the state
\begin{equation}\label{equation:state_vector_LTV_system}
    x=\begin{bmatrix}\posbfr\\ \velbfr\\ \gravbfr \\ \magbfr\end{bmatrix},
\end{equation}
we obtain, in view of \eqref{equation:dynamic_model_Body_frame} and \eqref{equation:virtual_nul_output}, an LTV system of the form:
\begin{subequations}\label{equation:LTV_state_model}
    \begin{align}
    \dot{x}&=A(t)x+B\accelbfr,\\
    y&=C(t)x,
    \end{align}
\end{subequations}
with matrices $A(t), B$ and $C(t)$ given by
\begin{align*}
A(t)&=\begin{bmatrix}-[\omega(t)]_\times & I_3 & \othree& \othree \\
                      \othree & -[\omega(t)]_\times & I_3& \othree\\
                      \othree & \othree & -[\omega(t)]_\times  & \othree\\  
                      \othree & \othree &  \othree& -[\omega(t)]_\times
\end{bmatrix}, \\
B&=\begin{bmatrix}
    \othree & I_3 & \othree &\othree
\end{bmatrix}^{\top},\\
C(t)&=\begin{bmatrix}
    \Pi_{\bearingbfr(t)} & \othree & \othree& \othree\\
    \othree & \othree & \othree& I_3
\end{bmatrix}.
\end{align*}
Note that matrix $A(t)$ is time-varying since it depends on the profile of the angular velocity $\omega(t)$ which can be seen as an external time-varying signal. Similarly, matrix $C(t)$ is also time-varying since it depends on the time-varying bearing  $\bearingbfr(t)$. Furthermore, we impose the following realistic constraint on the system's trajectory which is needed to ensure that the matrix $A(t)$ is well-conditioned for the convergence guarantees of the Riccati observer \eqref{equation:Riccati_observer}.
\begin{assumption}\label{assumption::bounded_bw}
The angular velocity is continuously differentiable and uniformly
bounded with bounded derivatives.
\end{assumption}
The state of system \eqref{equation:state_vector_LTV_system} is then estimated using the Riccati observer:
\begin{align}\label{equation:Riccati_observer_full_state}
\dot{\hat{x}}&=A(t)\hat{x}+B\accelbfr+K(t)(y-C(t)\hat{x}),
\end{align}
where $\hat x:=[\hat p^{\mathcal{B}}\;\hat v^{\mathcal{B}}\;\hat g^{\mathcal{B}}\;\hat m^{\mathcal{B}}]$ with $\hat p^{\mathcal{B}}$, $\hat v^{\mathcal{B}}$, $\hat g^{\mathcal{B}}$, $\hat m^{\mathcal{B}}$ are the estimates of $\posbfr$, $\velbfr$, $\gravbfr$, $\magbfr$, respectively, and $K(t)$ is computed using \eqref{equation:gain_general_Riccati_observer} and \eqref{equation:Riccati_equation}.
In the next Lemma, we analyse the uniform observability of the pair $(A(t),C(t))$ in the sense of Definition~\ref{definition:uniform_observability} which is necessary for the convergence of the observer. 
\begin{lemma}\label{Lemma:sufficient_conditon_for_uniform_observability}Let $\bearingifr:=R\bearingbfr$ be the bearing-to-landmark expressed in the inertial frame. If there exist $\delta,\mu>0$ such that
\begin{equation}\label{equation:conditon_lemma_1}
    \forall t\geq 0:\frac{1}{\delta}\int_t^{t+\delta}\Pi_{\bearingifr(s)} ds\geq\mu I_3,
\end{equation}
then the pair $(A(\cdot),C(\cdot))$ in \eqref{equation:LTV_state_model} is uniformly observable.
\end{lemma}
\begin{proofa}
Let us first compute the state transition matrix for \eqref{equation:LTV_state_model}. Let $\mu>0$ and define the following matrices:
\begin{align}
     T(t)&:= \begin{bmatrix}
     R(t) & \othree & \othree  & \othree \\
     \othree & R(t) & \othree & \othree  \\
     \othree & \othree & R(t) & \othree  \\
     \othree & \othree & \othree & R(t)
         \end{bmatrix},\nonumber\\ \bar{A}&:=\begin{bmatrix}
             \othree & I_3 & \othree & \othree\\
             \othree & \othree & I_3 & \othree\\
             \othree & \othree & \othree & \othree\\
             \othree & \othree & \othree & \othree
         \end{bmatrix}, \\
         \bar{C}&:=\begin{bmatrix}
             I_3& \othree &\othree &\othree \\
              \othree &\othree &\othree &\frac{1}{\sqrt{\mu}}I_3
         \end{bmatrix}.  \nonumber
\end{align}
Consider the change of variable $z(t)=T(t)x(t)$ and let $a^{\mathcal{B}}\equiv0$. Then, by direct differentiation one obtains $\dot z =\bar{A}z$ which implies that $z(t)=\exp(\bar{A}(t-s))z(s)$ for any $0\leq s \leq t$. Therefore, $x(t)=T(t)^{\top}\exp(\bar{A}(t-s))T(s)x(s)$ which implies that the state transition matrix is given by 
\begin{align}\label{equation:proof:transition_matrix}
\phi(t,s)&=T(t)^{\top}\exp(\bar{A}(t-s))T(s)\nonumber\\
&=:T(t)^{\top}\bar{\phi}(t,s)T(s). 
\end{align}
We now show that the observability Gramian of the pair $(A(t),C(t))$ satisfies \eqref{equation:conditon_of_uniform_observability}. The observability Gramian is written as:
\begin{equation}\label{equation:proof:grammian_matrix}
    W(t,t+\delta)=\frac{1}{\delta}\int_{t}^{t+\delta}\phi^{\top}(s,t)C^{\top}(s)C(s)\phi(s,t)ds.
\end{equation}
Now, in view of \eqref{equation:proof:transition_matrix}, \eqref{equation:proof:grammian_matrix} yields
\begin{flalign}\label{equation:proof:grammian_matrix_with_phi_bar}
\quad \quad    &W(t,t+\delta)=\\
\quad &\frac{1}{\delta}\int_{t}^{t+\delta}T^{\top}(t)\bar{\phi}^{\top}(s,t)T(s)C^{\top}(s)C(s)T(s)^{\top}\bar{\phi}(s,t)T(t)ds.\nonumber
\end{flalign}
Since  $C(s)=Q(s)\bar{C}$ with
\begin{equation}
   Q(s)=\begin{bmatrix}
      \Pi_{\bearingbfr(s)} & \othree \\
      \othree & \sqrt{\mu}I_3
    \end{bmatrix},
\end{equation}
we have
\begin{align}\label{equation:proof:grammian_matrix_with_M}
W(t,t+\delta)=
T^{\top}(t)\left(\frac{1}{\delta}\int_{t}^{t+\delta}\bar{\phi}^{\top}(s,t)M(s)\bar{\phi}(s,t)ds\right)T(t),
\end{align}
with $M(s):=T(s)\bar{C}^{\top}Q(s)Q(s)\bar{C}T(s)^{\top}$. Moreover, we have 
\begin{equation}
\bar{C}T(s)^{\top}=\begin{bmatrix}
    R(s)^{\top} & \othree \\
    \othree    & R(s)^{\top}
\end{bmatrix}\bar{C},
\end{equation}
Thus \eqref{equation:proof:grammian_matrix_with_M} becomes
\begin{flalign}\label{equation:proof:grammian_matrix_with_Sigma}
W&(t,t+\delta)=\nonumber\\
&T^{\top}(t)\left(\frac{1}{\delta}\int_{t}^{t+\delta}\bar{\phi}^{\top}(s,t)\bar{C}^{\top}\Sigma(s)\bar{C}\bar{\phi}(s,t)ds\right)T(t),
\end{flalign}
with 
\begin{equation}
\Sigma(s)=\begin{bmatrix}
    R(s) & \othree \\
    \othree    & R(s)
\end{bmatrix}Q(s)Q(s)\begin{bmatrix}
    R(s)^{\top} & \othree \\
    \othree    & R(s)^{\top}
\end{bmatrix}.   
\end{equation}
Since  $\Pi_{\bearingifr(s)}\Pi_{\bearingifr(s)}=\Pi_{\bearingifr(s)}$, we have
\begin{equation}
\Sigma(s)=\begin{bmatrix}
    \Pi_{\bearingifr(s)} & \othree \\
    \othree    &    \mu I_3    
\end{bmatrix}.
\end{equation}
Hence, in view of \eqref{equation:conditon_lemma_1}, there exist $\delta,\mu>0$ such that $\Sigma(s)$ satisfies
\begin{equation}\label{equation:conditon_lemma_1_with_vector_measurement}
    \forall t\geq 0:\frac{1}{\delta}\int_t^{t+\delta}\Sigma(s)ds\geq\mu I_6.
\end{equation}
Moreover, $\bar{A}$ and $\bar{C}$ are constant matrices, the pair $(\bar{A},\bar{C})$ is Kalman observable and $A$ has real eigenvalues. Therefore, it follows from \cite[Lemma 2.7]{Hamel2017PositionMeasurements}
that there exist $\bar{\delta}$ and $\bar{\epsilon}$ such that
\begin{flalign}\label{equation:proof:last_equation}
\frac{1}{\bar{\delta}}\int_{t}^{t+\bar{\delta}}\bar{\phi}^{\top}(s,t)\bar{C}^{\top}\Sigma(s)\bar{C}\bar{\phi}(s,t)ds\geq\bar{\epsilon} I_{12},
\end{flalign}
which also implies, in view of \eqref{equation:proof:grammian_matrix_with_Sigma} and since $T(t)^{\top}T(t)=I_{12}$, that $W(t, t+\bar{\delta})\geq \bar{\epsilon}I_{12}$ for all $t\geq0$. Therefore, the pair $(A(\cdot),C(\cdot))$ is uniformly observable, and the proof is complete.
\end{proofa}

Lemma~\ref{Lemma:sufficient_conditon_for_uniform_observability} provides a persistency of excitation condition on the matrix $\Pi_{\bearingifr}$. This PE condition is essentially equivalent to requiring that $|\dot \bearingifr|$ is regularly larger than a positive number \cite{Hamel2017PositionMeasurements}. In other words, this requires that the vehicle is never static nor indefinitely moving in a straight line with the landmark \cite{Berkane_Automatica_2021}.

Since the pair $(A(t), C(t))$ is uniformly observable under the PE condition of Lemma~\ref{Lemma:sufficient_conditon_for_uniform_observability}, we can design a Riccati observer \cite[Section 2.2]{Hamel2017PositionMeasurements} to estimate the position $\posbfr$, the velocity $\velbfr$, the gravity vector in the body frame $\gravbfr$ and the body-frame vector measurement $\magbfr$. Once good estimates of $\gravbfr$ and $\magbfr$ are available, the orientation matrix can be computed using algebraic reconstruction (see \cite[Corollary 6]{MARTIN201715409}) as follows: 
\begin{equation}\label{eq:hatR}
 \hat{R}^{\top}=\begin{bmatrix}
\frac{\hatgravbfr}{|\gravifr|} &  \frac{\hatgravbfr\times \hatmagbfr}{|\gravifr\times\magifr|} & \frac{\hatgravbfr\times(\hatgravbfr\times\hatmagbfr)}{|\gravifr\times(\gravifr\times\magifr)|}  
 \end{bmatrix}  \bar{R}^{\top}, 
\end{equation}
where
\begin{equation}
 \bar{R}:=\begin{bmatrix}
\frac{\gravifr}{|\gravifr|} &  \frac{\gravifr\times \magifr}{|\gravifr\times\magifr|} & \frac{\gravifr\times(\gravifr\times\magifr)}{|\gravifr\times(\gravifr\times\magifr)|}   
 \end{bmatrix}.
\end{equation}
It is important to note that $\hat{R}$ as defined above is not necessary a rotation matrix but converges to a rotation matrix. However, if it is required to work with a rotation matrix at all time, a simple solution is to project $\hat{R}$ to the nearest rotation matrix using polar decomposition as explained in \cite[Proposition 7]{MARTIN201715409}. Note that \eqref{eq:hatR} is implementable under Assumption \ref{assumption::obsv}. 
\begin{remark}[Attitude Estimation on $\so$]~An~alternative solution to estimate the full attitude matrix $\hat R$ is to cascade the proposed linear observer with a nonlinear complementary filter on $\so$ such as in \cite{mahony2008nonlinear,berkane2017design}. Thanks to the proposed observer being globally exponentially convergent and the almost global input-to-state stability (ISS) property of the nonlinear complementary filters on $\so$ (see \cite{wang2021nonlinear}), it is not difficult to show that the interconnection preserves almost global asymptotic stability of the estimation errors. 
\end{remark}
\begin{remark}[Decoupled Observer]\label{remark2}
Note that, in practice, the estimation of the pitch and roll angles independently from the magnetic disturbances holds significant importance for ensuring robust flights for UAVs \cite[Section~\RNum{2}-C]{MinhDuc_TCST_2013}. In fact, thanks to the structure of the system's matrices and by choosing the   parameters of the Riccati equation as follows:
\begin{itemize}
    \item The initial condition $P(0)=\mathrm{blockdiag}(P_1(0),P_2(0))$ with $P_1(0)\in\R^{9\times 9}$ and $P_2(0)\in\R^{3\times 3}$ as positive definite matrices,
    \item The matrices $V(t)=\mathrm{blockdiag}(V_1(t),V_2(t))$ and $Q(t)=\mathrm{blockdiag}(Q_1(t),Q_2(t))$, where $V_1(t)\in\R^{9\times 9}$  and  $V_2(t),  Q_1(t), Q_2(t)\in\R^{3\times 3}$ as uniformly positive definite matrices,
\end{itemize}
the obtained observer's structure becomes decoupled and its gain will be written, in view of \eqref{equation:gain_general_Riccati_observer}  and \eqref{equation:LTV_state_model}, as follows:
\begin{equation}\label{equation:gain_decoupled_observer}
K(t)=\begin{bmatrix}
    P_1(t)C_1(t)^{\top}Q_1(t) & \othree \\
    \othree                       & P_2(t)Q_2(t)
\end{bmatrix},
\end{equation}
with $C_1(t)=\begin{bmatrix}
\Pi_{\bearingbfr} & \othree & \othree
\end{bmatrix}$ and the estimates of $\posbfr$, $\velbfr$ and $\gravbfr$ will be independent of the estimates of $\magbfr$. Thus, the pitch and roll estimates can be extracted from  $\hatgravbfr$ independently from the estimates of yaw, which can be obtained from $\hatmagbfr$.
\end{remark}
\begin{remark}[Reduced-Order Observer]
An alternative solution to decouple the roll/pitch estimation from yaw estimation would be to consider the following reduced state $\tilde{x}=\begin{bmatrix}
 \posbfr & \velbfr &\gravbfr   
\end{bmatrix}^{\top}$. Then, if the condition of Lemma~\ref{Lemma:sufficient_conditon_for_uniform_observability} is satisfied, the pair $(\tilde{A}(t),\tilde{C}(t))$ given by \begin{align}
    \tilde{A}(t)&=\begin{bmatrix}-[\omega(t)]_\times & I_3 & \othree \\
                      \othree & -[\omega(t)]_\times & I_3\\
                      \othree & \othree & -[\omega(t)]_\times  \\  
\end{bmatrix},\nonumber\\ 
\tilde{C}(t)&=\begin{bmatrix}
    \Pi_{\bearingbfr} & \othree & \othree
\end{bmatrix}\nonumber
\end{align}
is uniformly observable and the estimates of $\tilde{x}$  is computed using the Riccati observer:
\begin{align}\label{equation:Riccati_observer_reduced_state}
\dot{\hat{\tilde{x}}}&=\tilde{A}(t)\hat{\tilde{x}}+\tilde{B}\accelbfr+\tilde{K}(t)(\tilde{y}-\tilde{C}(t)\hat{\tilde{x}}),
\end{align}
    with $\hat{\tilde{x}}=\begin{bmatrix}
    \hat{p}^{\mathcal{B}} & \hat{v}^{\mathcal{B}} & \hat{g}^{\mathcal{B}}\end{bmatrix}$, $\bar{B}=\begin{bmatrix}
    \othree & I_3 & \othree 
\end{bmatrix}^{\top}$, $\tilde{y}$ is given in \eqref{equation:virtual_nul_output} and $\tilde{K}(t)$ is computed similarly to \eqref{equation:gain_general_Riccati_observer} and \eqref{equation:Riccati_equation}.
On the other hand, the pitch and roll angles can be computed from the estimates of $\gravbfr$. In fact, if we consider the ZYX convention, the last row of the rotation matrix $e_3^\top R$ is given by \cite{henderson1977euler}
\begin{align}
    e_3^\top R=\begin{bmatrix}
        -\sin(\theta)&\cos(\theta)\sin(\phi)&\cos(\theta)\cos(\phi)
    \end{bmatrix},
\end{align}
where $\theta$ and $\phi$ are the pitch and roll, respectively. Now, since $\gravifr=g^{\mathcal{I}}_3e_3$, one has $\gravbfr/g^{\mathcal{I}}_3=R^\top e_3$ and, therefore, the pitch and roll are given by
\begin{align} 
\theta&=\mathrm{atan2}\left(-\gravbfr_1,\sqrt{(\gravbfr_2)^2+(\gravbfr_3)^2}\right),\\
\phi&=\mathrm{atan2}(\gravbfr_2,\gravbfr_3).
 \end{align}
Therefore, since only $\hatgravbfr$ is available, the estimated roll and pitch are obtained using:
\begin{align} 
\hat{\theta}&=\mathrm{atan2}\left(-\hatgravbfr_1,\sqrt{(\hatgravbfr_2)^2+(\hatgravbfr_3)^2}\right),\\
 \hat{\phi}&=\mathrm{atan2}(\hatgravbfr_2,\hatgravbfr_3).
 \end{align}
Yaw estimates can be provided independently using external sources such as dual-GPS or compasses. 
\end{remark}
\section{Simulation Results}\label{section:simulation}
In this section, we obtain simulation results to test the performance of the observer proposed in Section~\ref{Section:main_result}.
Consider a vehicle moving in 3D space and tracking the following eight-shaped trajectory:
\begin{equation}\label{equation:trajectory_simulation}
p(t)=\begin{bmatrix}\cos(5t) \\ \sin(10t)/4 \\ -\sqrt{3}\sin(10t)/4]\end{bmatrix}.
\end{equation}
The rotational motion of the vehicle is subject to the following angular velocity:
\begin{equation}
 \omega(t)=\begin{bmatrix}\sin(0.1t+\pi) \\ 0.5\sin(0.2t) \\ 0.1\sin(0.3t+\pi/3)\end{bmatrix}.
\end{equation}
The initial values of the true pose are $\posif(0)=\begin{bmatrix}
1 & 0 &0
\end{bmatrix}^{\top}$, $\velif(0)=\begin{bmatrix}
    -0.0125 & 2.5 & -4.33
\end{bmatrix}^{\top}$ and $R(0)=\exp([\pi e_2]_{\times}/2)$ with $e_2=\begin{bmatrix}
    0 &1& 0
\end{bmatrix}^{\top}$. The initial conditions for the observer are $\hat{x}(0)=\begin{bmatrix}1& 1& 1& 1& 1& 1& 4.9& 4.9& 4.9\end{bmatrix}^{\top}$,
$P(0)=I_9$, $V(t)=36I_9$, $Q(t)=I_3$ and $\hat{R}(0)=I_3$. The gravity vector $\gravifr$ is set to $\begin{bmatrix}
 0 & 0 & 9.81   
\end{bmatrix}^{\top}$ while the constant vector $\magifr$ is set to $[\begin{matrix}
 \tfrac{1}{\sqrt{2}} & 0 & \tfrac{1}{\sqrt{2}}   
\end{matrix}]^{\top}$, which mimics the magnetic filed. The body-frame vector measurement $\magbfr$ is considered to be affected by Gaussian white noise with noise-power $10^{-2}$. The results are presented in Figs.~\ref{figure:real_and_estimated_position}-\ref{figure:real_and_filtered_vector_measurement}. One can easily observe that the estimate trajectories converge to the real ones after some seconds. In overall, we find the performance of the proposed observer quite satisfactory.  Note that the considered trajectory \eqref{equation:trajectory_simulation} is rich enough to satisfy the PE condition of Lemma~\ref{Lemma:sufficient_conditon_for_uniform_observability} and ensures exponential convergence of the observer. 

\begin{figure}[h!]
    \centering
\includegraphics[scale=0.6]{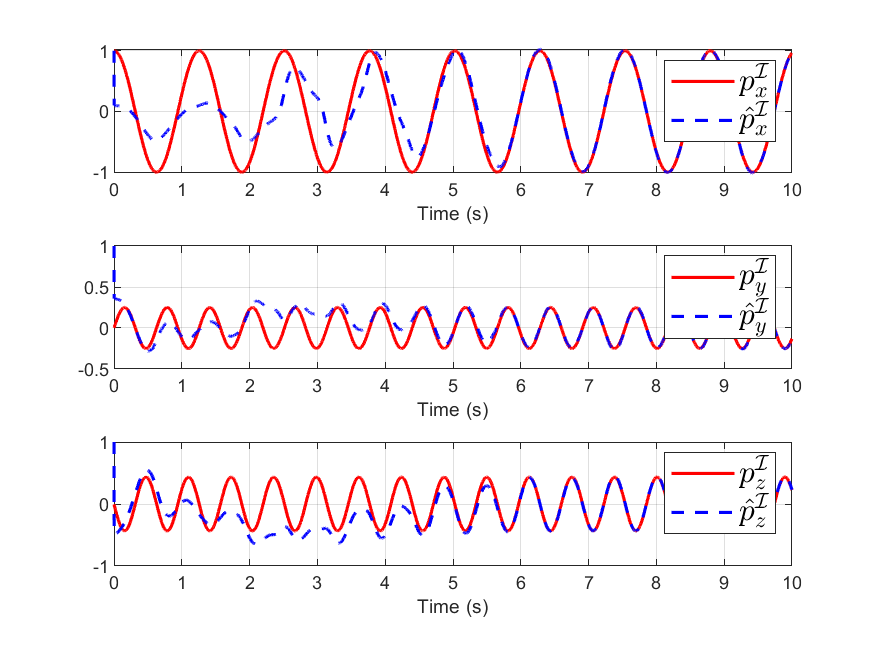}
    \caption{Time behaviour of the components of the real and
estimated position.}
    \label{figure:real_and_estimated_position}
\end{figure}

\begin{figure}[h!]
    \centering
\includegraphics[scale=0.6]{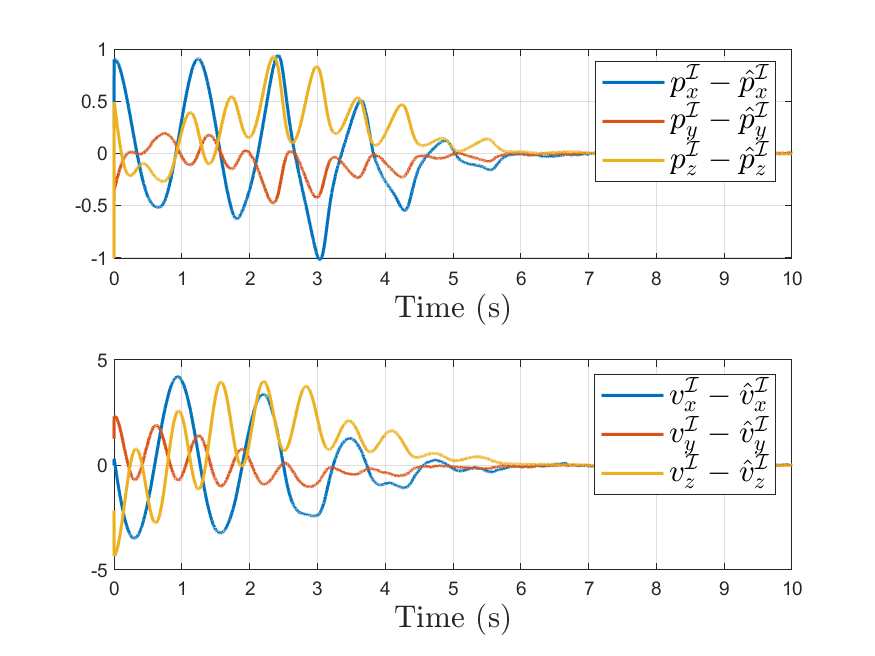}
    \caption{Position and velocity estimation errors.}
    \label{figure:estimation_error}
\end{figure}
\begin{figure}
    \centering
\includegraphics[scale=0.63]{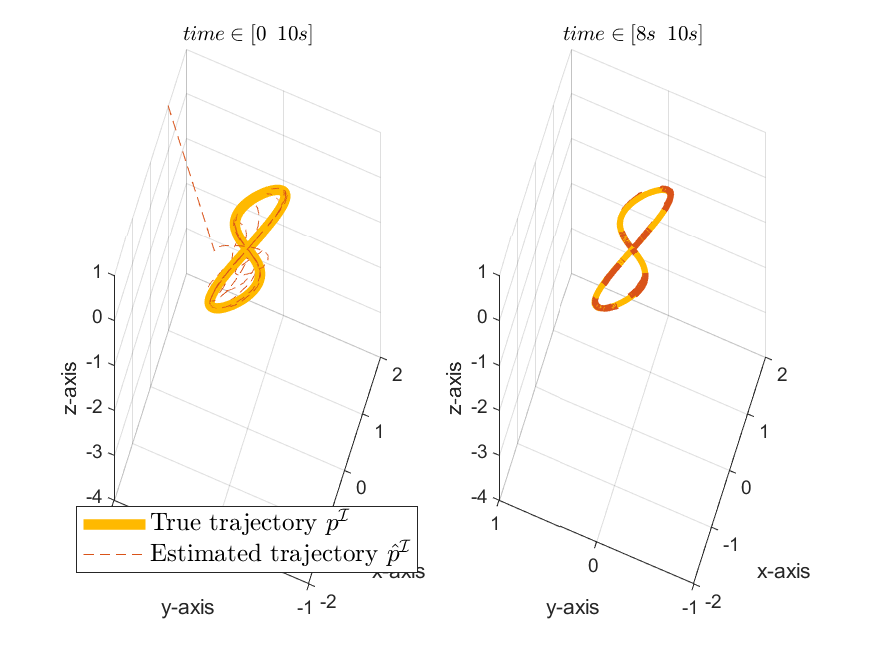}
    \caption{True and estimated trajectory in the inertial frame.}
    \label{figure:3D_real_and_estimated_position}
\end{figure}

\begin{figure}[h!]
    \centering
\includegraphics[scale=0.62]{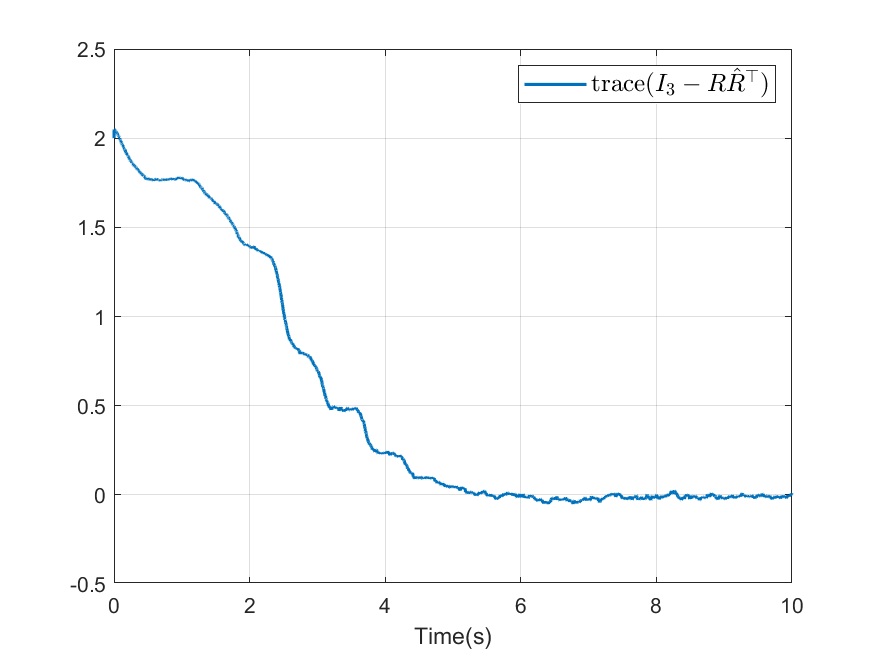}
\caption{Attitude estimation error.}
    \label{figure:estimation_error_of_the_orientation}
\end{figure}

\begin{figure}[h!]
    \centering
\includegraphics[scale=0.57]{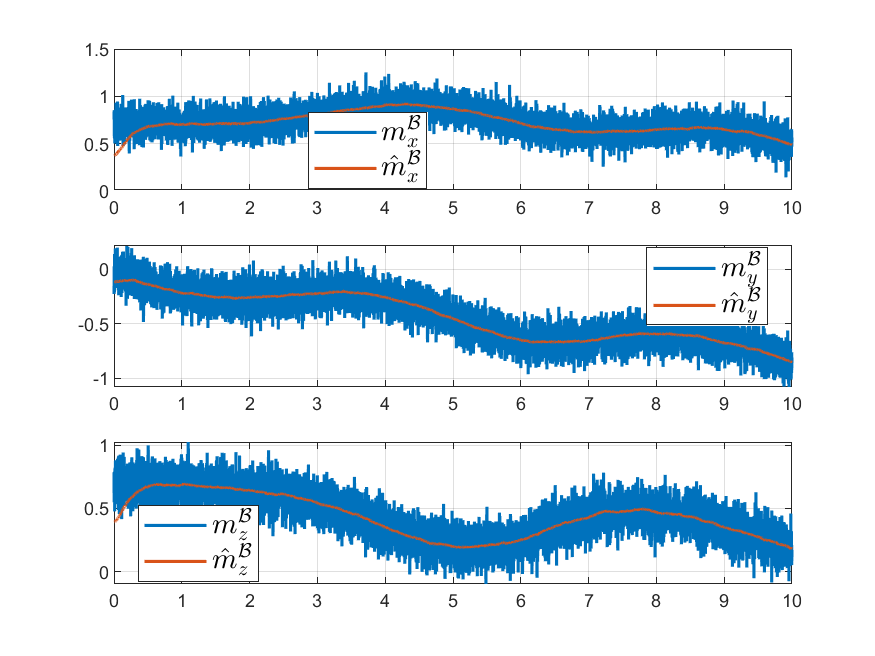}
    \caption{Components of the measured and estimated (filtered) magnetic field vector expressed in the body frame.}
    \label{figure:real_and_filtered_vector_measurement}
\end{figure}

\section{Conclusion}\label{section:conclusion}
In this work, we proposed a Riccati-based observer for simultaneous position, linear velocity and attitude estimation of a rigid body. The proposed observer uses measurements from IMU (acceleration in the body frame and angular velocity), a single bearing measurement and a body-frame vector observation. A detailed uniform observability analysis (UO) has been curried out and sufficient conditions for UO are derived as a persistency of excitation (PE) condition on the trajectory. This PE condition guarantees global exponential convergence of the proposed estimator. Furthermore, under the same PE condition, a reduced-order form of the estimator has been discussed in Remark \ref{remark2} which allows to estimate body-frame position, velocity, and gravity using only IMU and single bearing. This allows to estimate roll and pitch independently from the other body-frame vector observation (\textit{e.g.,} magnetometer). As a future work, we intend to improve our proposed approach by considering biased IMU measurements while preserving the global convergence property.
\bibliographystyle{unsrt}
\bibliography{main}

\end{document}